\documentclass[twocolumn,showpacs,preprintnumbers,amsmath,amssymb,floatfix]{revtex4}
\usepackage{graphicx}% Include figure files
\usepackage{dcolumn}% Align table columns on decimal point
\usepackage{bm}% bold math

\begin{document}

\preprint{APS/123-QED}

\title{
An exact study of charge-spin separation, pairing  fluctuations \\
and pseudogaps in
four-site Hubbard clusters}

\author{Armen~N.~Kocharian} \email{armen.n.kocharian@csun.edu}

\affiliation{%
Department of Physics and Astronomy, California State University,
Northridge, \\ CA 91330-8268
}%

\author{Gayanath~W.~Fernando}
\email{fernando@phys.uconn.edu}
\affiliation{Department of Physics, University of Connecticut,\\
 Storrs, CT 06269 and IFS, Hantana Rd., Kandy, Sri Lanka
}%

\author{Kalum~Palandage}
\affiliation{Department of Physics, University of Connecticut,\\
 Storrs, CT 06269
}%

\author{James~W.~Davenport}
\affiliation{Computational Science Center, Brookhaven National
Laboratory,\\
Upton, NY 11973}

\begin{abstract}

An exact study of charge-spin separation, pairing fluctuations and
pseudogaps is carried out by combining the {\it analytical}
eigenvalues of the four-site Hubbard clusters with the grand
canonical and canonical ensemble approaches in a multidimensional
parameter space of temperature ($T$), magnetic field ($h$),
on-site interaction ($U$) and chemical potential ($\mu$). Our
results, near the average number of electrons $\left\langle
N\right\rangle\approx 3$, strongly suggest the existence of a
critical parameter $U_{c}(T)$ for the localization of electrons
and a particle-hole binding ({\sl positive}) gap
$\Delta^{e-h}(T)>0$ at $U>U_{c}(T)$, with a zero temperature
%critical value,
quantum critical point, $U_{c}(0)=4.584$.
For $U<U_{c}(T)$, 
particle-particle pair binding is found with a ({\sl positive})
pairing gap
$\Delta^{P}(T)>0$.
The ground state degeneracy is lifted at $U>U_c(T)$ and the
cluster becomes a Mott-Hubbard like insulator due to the presence
of energy gaps at all (allowed) integer numbers ($1\le N\le 8$) of
electrons. In contrast, for $U\le U_c(T)$, we find an electron
pair binding instability at finite temperature near $\left\langle
N\right\rangle\approx 3$, which manifests a possible pairing
mechanism, a precursor to superconductivity in small clusters.
Rigorous criteria for the existence of many-body Mott-Hubbard like
particle-hole and particle-particle pairings,
%spin
spin-spin
pairing, (spin) pseudogap and (spin) antiferromagnetic
critical crossover temperatures, at which the corresponding
pseudogaps disappear, are also formulated. In particular, the
resulting phase diagram consisting of charge and spin pseudogaps,
antiferromagnetic correlations, hole pairing with competing
hole-rich ($\left\langle N\right\rangle=2$), hole-poor
($\left\langle N\right\rangle=4$) and magnetic ($\left\langle
N\right\rangle=3$) regions in the ensemble of clusters near 1/8
filling closely resembles the phase diagrams and inhomogeneous
phase separation recently found in the family of doped high T$_c$
cuprates.

\end{abstract}
                             % Classification Scheme.

\maketitle

\section{Introduction}
Understanding the effects of electron correlation and pseudogap
phenomena
~\cite{RVB,Nature,Timusk,Marshall,Kivelson_Review,Andrea} in doped
oxides, including the cuprate superconductors is one of the most
challenging problems in condensed matter physics~\cite{Anderson}.
Although the experimental determination of various inhomogeneous
phases in cuprates is still somewhat controversial~\cite{Tallon},
the underdoped high  T$_c$ superconductors (HTSCs) are often
characterized by crossover temperatures below which excitation
pseudogaps in the normal-state are seen to develop~\cite{Zachar}.
In these materials, the spectral weight begins to be strongly
suppressed below some characteristic temperature  T$_s$ that
is higher than the superconducting crossover temperature
T$_c$. There are many experiments supporting a highly nonuniform
hole distribution leading to the formation of hole-rich and
hole-poor regions in doped La$_{2-x}$Sr$_x$CuO$_4$ and other
cuprate HTSCs ~\cite{Matsuda}. This electronic phase separation is
expected to be mostly pronounced at low hole concentrations.
Recently, strong experimental evidence has been found for
`electronic phase separation'  in La-cuprates near optimal doping
into separate, magnetic and superconducting phases~\cite{Hashini}.

The relevance of the Hubbard model for studies of the HTSCs has
been the focus of intensive research and  debated for quite some
time with no firm conclusions up to now. Even though the small
Hubbard clusters do not have the full capacity to describe the
complexity of copper ions and their ancillary oxygens detected in
the HTSC materials, it is still argued that this model can capture
the essential physics of the HTSCs~\cite{RVB}. However, beyond one and
infinite dimensions, there is no exact solution currently available
for the Hubbard Hamiltonian. It is also known that in the
optimally doped cuprates, the correlation length of dynamical spin
fluctuations is very small~\cite{Zha}, which points to the local
character of electron interactions in the cuprates. Therefore, exact
microscopic studies of pairing, crossover and pseudogaps, by using
analytical diagonalization of small Hubbard clusters, which
account accurately for short-range dynamical correlations, are
relevant and useful with regard to understanding the physics of
the HTSCs. Our exact analytical solution appears to be providing
useful insight into the physical origin of the high energy
insulator-metal and low energy antiferromagnetic crossovers,
electron pairing and spin density fluctuations in the
superconducting phase.

The following questions are central to our study: (i) When treated
exactly, what essential features can the simple Hubbard clusters
capture, that are in common with the HTSCs? (ii) Using simple
cluster studies, is it possible to obtain a mesoscopic
understanding of electron-hole/electron-electron pairing and
identify various possible phases and crossover temperatures? iii)
Do these small clusters (coupled to a particle bath) contain
important features that are similar to large clusters and
thermodynamical systems?

Our work has uncovered important answers to the questions raised
above. This is a follow-up to our recent study,
 in which rigorous criteria were
found for the existence of
microscopic quantum critical points (QCP),
Mott-Hubbard (MH) and Ne{\'e}l type bifurcations,
and
corresponding critical temperatures of crossovers and various
phases in finite-size systems ~\cite{JMMM}. Small 2- and 4-site
clusters with short range electronic correlations provide (unique)
insight into the thermodynamics and exact many body physics,
difficult to obtain from approximate methods. In particular, we
show that these small Hubbard clusters, in the absence of long
range order, exhibit  particle-particle, Mott-Hubbard like
particle-hole or antiferromagnetic-paramagnetic instabilities
in the ground state and
at finite
%temperature.
temperatures.

%%%%%%%%%%%%%%%%%%%%%%%
In addition, the 4-site (square) cluster is the basic building
block of the CuO$_2$ planes in the HTSCs and it can be used as a
block reference to build up larger superblocks in 2D of desirable
sizes by applying Cluster Perturbation  Theory (CPT)~\cite{CPT},
non-perturbative Real-Space Renormalization Group
(RSRG)~\cite{rsrg}, Contractor Renormalization Group
(CORE)~\cite{CRG}, or Dynamical Cluster Approximation (DCA) for
embedded 4-site clusters coupled to an uncorrelated
bath~\cite{Jarrell}. Similar attempts at studying small clusters
(such as the ground state studies of weakly coupled Hubbard dimers
and squares by Tsai and Kivelson~\cite{tk}) have begun recently
and the lessons learned here will be invaluable to such studies
and useful to the condensed matter community in general. Above
all, these small clusters can be synthesized and utilized for
understanding essential many-body physics at the mesoscopic level
and hence are useful in their own right.

\section{Model and methodology}

The single orbital 4-site minimal Hubbard Hamiltonian,
\begin{eqnarray}
 H = -t\sum\limits_{i \sigma}(c^{+}_{i\sigma}c_{i+1\sigma}
+ H.c.)+U\sum\limits_{i}n_{i\uparrow}
n_{i\downarrow}-h\sum\limits_{i}S^z_i,\label{2-site1}
\end{eqnarray}
with hopping $t$ and on-site interaction $U$ is the focus of this
work. Periodic boundary conditions are used for the clusters.
Unless otherwise stated, all the energies reported here are
measured in units of $t$ (i.e. $t$ has been set to $1$ in most of
the equations that follow). Our calculations are based on the
exact {\sl analytical} expressions for the eigenvalue $E_{n}$ of
the $n^{th}$ many-body eigenstate of the 4-site Hubbard
clusters~\cite{schumann}. As we show here, this model, used in
conjunction with the grand canonical and canonical ensembles,
yields valuable insight into the physics of strongly correlated
electrons.

\begin{figure} %[h]
\begin{center}
\includegraphics*[width=20pc,height=20pc]{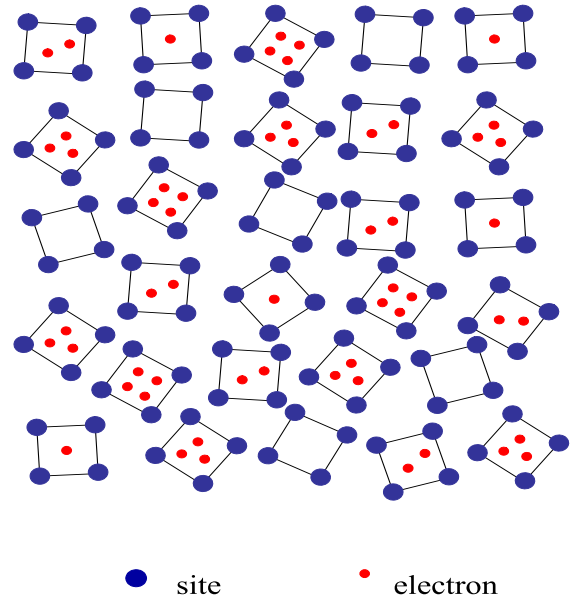}
\end{center}
\caption {Various possible configuration mixing of electrons (below half
filling) that can be found in an ensemble of 4-site clusters. Note
that the mixing of configurations is brought about by the
temperature.}
 \label{fig:ensemble}
\end{figure}

\subsection{
Thermodynamics and response functions} The complete phase diagram
of interacting electrons can be obtained with high accuracy due to
the analytically (exact) given thermodynamic expressions. In
Fig.~\ref{fig:ensemble}, possible electron configurations (below
half-filling) in the grand canonical ensemble for the 4-site
clusters are shown. The grand partition function Z$_U$ (where the
number of particles $N$ and the projection of spin $s^z$ are
allowed to fluctuate) and its derivatives are calculated (exactly)
without taking the thermodynamic limit. The exact grand canonical
potential $\Omega_U$ for many-body interacting electrons is
\begin{eqnarray}
\Omega_U=-{T}\ln \!\sum\limits_{n\leq N_H} e^{{-\frac {E_{n}-\mu
N_{n} - hs^{z}_n} T} }, \label{OmegaU}
\end{eqnarray}
where $N_n$ and $s^{z}_n$ are the number of particles and the
projection of spin in the $n^{th}$ state respectively. The Hilbert
space dimension in (\ref{OmegaU}) is $N_H=4^4$. The derivatives we
study may be labeled as
 first order (such as the average spin projection/magnetization
 in response to an applied magnetic field) or second order (such as fluctuations/susceptibilities).
 These responses, evaluated as functions of chemical potential $\mu$,
 applied field $h$, on-site Coulomb interaction $U$ and temperature $T$,
 carry a wealth of information that can be used
 to identify various phases and phase boundaries.
 Some of these results for the 2- and 4-site clusters
 were reported earlier ~\cite{JMMM}.

The  (first order) responses due to
doping and external magnetic field are as follows:
\begin{eqnarray}
\left\langle N\right\rangle=-{{\frac {\partial \Omega_U} {\partial
\mu}}}, \label{number}
\end{eqnarray}
\begin{eqnarray}
\left\langle s^z\right\rangle=-{{\frac {\partial \Omega_U} {\partial
h}}}. \label{spin}
\end{eqnarray}
Analytical expressions derived for the averages $\left\langle N\right\rangle$
and $\left\langle s^z\right\rangle$  are analyzed
numerically in a wide range of $U$, $h$, $\mu$ and $T$ parameters.
 The charge and spin degrees respond to an applied
magnetic field ($h$) as well as electron or hole doping levels
(i.e. chemical potential $\mu$) and display clearly identifiable,
prominent peaks, paving the way for rigorous definitions of
Mott-Hubbard (MH), antiferromagnetic (AF), spin pseudogaps and
related crossover behavior~\cite{JMMM}. The exact expressions for
charge susceptibility, ${\chi_c}={{\frac {\partial
{\left\langle {N}\right\rangle}} {\partial \mu}}}$ and spin
susceptibility, $\chi_s={{\frac {\partial \left\langle
s^{z}\right\rangle} {\partial h}}}$ can be found as a function of
 $U$, $h$, $\mu$ and $T$ from,

\begin{eqnarray}
\left\langle X^2\right\rangle-\left\langle
X\right\rangle^2=T{{\frac {\partial {\left\langle
{X}\right\rangle}} {\partial x}}}, \label{fluctuation_number}
\end{eqnarray}
where $X$ corresponds to $N$ or $s^z$ and $x$ to $\mu$ or $h$.
Using maxima and minima in spin and charge susceptibilities, phase
diagrams in a $T$ {\it vs} $\mu$ plane for any $U$ and $h$ can be
constructed. This approach also allows us to obtain
%quantum critical points
QCP
and rigorous criteria for various transitions, such as the MH
crossover at half-filling and MH like bifurcations, using the
evolution of peaks in charge or spin susceptibility~\cite{JMMM}
(see below).

\subsection{Charge (pseudo) gap}

We define  canonical energies $\mu_{\pm}$,
\begin{eqnarray}
&\mu_{+}=E(M+1,M^\prime;U:T)-E(M,M^\prime;U,T)\label{mupm1}  \\
&\mu_{-}=E(M,M^\prime;U:T)-E(M-1,M^\prime;U:T)
\label{mupm2}
\end{eqnarray}
where $E(M,M^\prime;U:T)$ is the canonical (ensemble) energy with
a given number of electrons $N=M+M^{\prime}$ determined by the
number of up ($M$) and down ($M^\prime$) spins. At zero
temperature the expressions (\ref{mupm1}) and (\ref{mupm2}) are
identical to those introduced in~\cite{Lieb1}. At finite
temperature, the calculated charge susceptibility is a
differentiable function of $N$ and $\mu$. The peaks (i.e. maxima)
in $\chi_c(\mu)$, {\sl which may exist in a limited range of
temperature}, are identified easily from the conditions,
$\chi^{'}_c(\mu_{\pm})=0$ with $\chi^{''}_c(\mu_{\pm}) <0$. We
define $T_c(\mu)$ to be the temperature at which a (possible) peak
is found in $\chi_c(\mu)$ at a given $\mu$.

The positive charge gap for electron-hole (excito) excitations,
${\Delta^{e-h}}(T)>0$, is defined by
\begin{eqnarray}
\Delta^{e-h}(T) = \left\{ \begin{array}{ll}
\mu_{+}-\mu_{-} & \mbox{if $\mu_{+} > \mu_{-}$} \\
  0 & \mbox{otherwise,} \\
\end{array}
\right.
\label{charge_gap}
\end{eqnarray}
as the separation between  $\mu_{+}$ and $\mu_{-}$. The
electron-hole instability
%$\Delta^{e-h}(T)\ge 0$
$\Delta^{e-h}(T)>0$
can exist in a limited range of temperatures and
$U>U_c(T)$,
with $\Delta^{e-h}(T)\equiv 0$ at $U\leq U_c(T)$. (In general, the
critical parameter $U_c(T)$, identified and discussed in
Sections~\ref{A} and \ref{B}, depends also on $h$.) The energy
gap, $\Delta^{e-h}(T)\ge 0$, serves as a natural order parameter
in a multidimensional parameter space ${h,U,T}$ and at $T\neq 0$
will be called a {\sl pseudogap}, since $\chi$ has a small, but
nonzero weight inside the gap. At $T=0$, this gap $\Delta^{e-h}_0
\equiv \Delta^{e-h}(0)$ will be labeled a {\sl true gap} since
$\chi_c$ is exactly zero inside.

The difference $\mu_{+}-\mu_{-}$ is somewhat similar to the
difference $I-A$ for a cluster, where $I$ is the ionization
potential and $A$ the electron affinity. For a single isolated
atom at half-filling and $T=0$, $I-A$ is equal to $U$ and hence
the above difference represents a screened $U$, reminiscent of
Herring's definition of $U$ in transition elements~\cite{herring}.

\subsection{Mott-Hubbard crossover}

The thermodynamic quantities with fixed $N$ ({\it canonical
approach}) are certainly smooth, analytical functions of $T$ and
$U$. Thus, one may naively think that at half-filling
and large $U$,
there is no real cooperative phenomena at $T\sim U$ for transition
from localized to delocalized electrons or at
$T\sim {t^2\over U}$
for transition from
antiferromagnetic to paramagnetic state. At finite temperature,
the thermodynamic quantities $\Omega_U$ and $\left\langle
N\right\rangle$ are both analytic and smooth functions of $T$ and
$\mu$. Although the charge susceptibility $\chi_c$ is also a
differentiable function of $\mu$ and $T$ at all $T>0$, $\chi_c$
{\it vs} $\mu$ exhibits a weak, fourth order singularity at some
critical temperature $T_{MH}$ ({\it saddle point})~\cite{JMMM}.
Thus the MH crossover at half-filling ($\mu = U/2$)
 can be defined simply
as a critical temperature $T_{MH}$, at which two peaks merge into
one with $\mu= \mu_{+} =\mu_{-}= U/2$ and $\chi^{'}_c(\mu) =0$ and
$\chi^{''}_c(\mu) =0$, i.e. as the temperature corresponding to a
point of inflexion in $\chi_c(\mu)$ ~\cite{JMMM}. This procedure
gives a rigorous definition for the MH crossover temperature
$T_{MH}$ (from a localized into an itinerant state), at which the
electron-hole pseudogap melts, i.e. $\Delta^{e-h}(T_{MH})=0$. The
MH crossover, due to its many-body nature, is also a cooperative
effect which may occur even for a single atom.

\subsection{Spin (pseudo) gap}

Analogously, we define a (positive) spin pairing gap  between
various spin configurations (projections of spin $s$) for a given
number of electrons ($N= M + M^\prime$)
in
the absence of a
 magnetic field ($h=0$) as,
\begin{eqnarray}
\Delta^s(T)=E(M+1,M^\prime-1;U:T)-E(M,M^\prime;U:T).
\label{spin_gap}
\end{eqnarray}
This corresponds to the energy necessary to make an excitation by
overturning a single spin. Possible peaks in the zero magnetic
field spin susceptibility $\chi_s(\mu)$, when monitored as a
function of $\mu$, can also be used to define an associated
temperature, $T_s(\mu)$, as the temperature at which such a peak
exists and the  spin pseudogap as the separation
(distance)
between two such peaks.

\subsection{AF (pseudo) gap and onset of magnetization}

Similar to the (charge) plateaus seen in $\left\langle
N\right\rangle$ {\it vs} $\mu$, we can trace the variation of
magnetization $\left\langle s^z\right\rangle$ {\it vs} an
applied magnetic field $h$ and identify the spin plateau features,
which can be associated with staggered magnetization or short
range antiferromagnetism. We calculate the critical magnetic field
$h_{c\pm}$ for the onset of magnetization ($s^z\to \pm 0$), which
depends on $N$ and $\mu$, by flipping a down spin,
$h_{c+}=E(M+1,M^\prime-1;U:T)-E(M,M^\prime;U:T)$ or an up spin,
$h_{c-}=E(M,M^\prime;U:T)-E(M-1,M^\prime+1;U:T)$~\cite{Sebold}.
The spin singlet binding energy ${\Delta^{AF}}(T)>0$ can be
defined as,
\begin{eqnarray}
{\Delta^{AF}}(T) = {{h_{c+}} - {h_{c-}}\over 2},
\label{af_gap}
\end{eqnarray}
and serves as a natural antiferromagnetic order parameter in a
multidimensional parameter space ${U,T,\mu}$. This will be
called an {\sl AF pseudogap} at nonzero temperature. We define
$T_{AF}$ as the temperature at which the pseudogap, $\Delta^{AF}(T)=0$,
vanishes and above which a paramagnetic state is found. An exact
analytical expression for the AF spin gap in the ground state
 ($\Delta^{AF}(0)$) at
half-filling was obtained in Ref.~\cite{JMMM}.
 In what follows, all of the temperatures defined above, $T_c(\mu)$,
$T_s(\mu)$ and $T_{AF}(\mu)$,
 will be used when constructing phase diagrams.

\subsection{Pairing instability}

To determine whether the cluster can support electron pairing at
finite temperature  despite the  purely repulsive electronic
interactions, the electron-electron (or hole-hole) pair binding
energy,
 \begin{eqnarray}
&\Delta^P(T)=
\nonumber \\
&[E(M-1,M^\prime;U:T)-E(M+1,M^\prime;U:T)]-
\nonumber \\
& 2[E(M,M^\prime;U:T)-E(M+1,M^\prime;U:T)],~
\label{pairing_gap}
\end{eqnarray}
is calculated by adding or subtracting one electron near
$N=M+M^\prime$. The average energy $E(M,M^\prime;U:T)$ is given
for configurations with a fixed number of electrons $N$ and spin
magnetization $s^z=0$ using our grand canonical ensemble approach.
At zero temperature, the binding energy (\ref{pairing_gap}) is
identical to the one introduced in Ref.~\cite{scalettar}.

Using the definitions for  $\mu_{\pm}$ from Eqs.~\ref{mupm1} and
\ref{mupm2}, electron-electron (or hole-hole) pair energy can also
be written as,

\begin{eqnarray}
\Delta^{P}(T) = \left\{ \begin{array}{ll}
\mu_{-}-\mu_{+} & \mbox{if $\mu_{-} > \mu_{+}$} \\
  0 & \mbox{otherwise.} \\
\end{array}
\right.
\label{pair_gap}
\end{eqnarray}
%%%%%%%%%%%%%%%%%%%%%%%%%%%%
In the ground state, the electron-pair binding energy gap
at $\left\langle N\right\rangle\approx 3$
is fully developed
at $U\leq U_c(T)$
when
$\Delta^P(0)>0$,
i.e. $\mu_{-}>\mu_{+}$, which leads to
the phase separation into $\left\langle N\right\rangle=2$ and
$\left\langle N\right\rangle=4$ clusters (see Section \ref{C})
and an effective attraction between the electrons
in $\left\langle N\right\rangle=2$ cluster configuration
%
%for $U\leq U_c(T)$
~\cite{white}. {\sl On the other hand, when $\mu_{+} > \mu_{-}$,
the condition ${\Delta^{e-h}}(T)>0$ with $U>U_c(T)$ provides an
electron-hole pairing mechanism as a precursor to
antiferromagnetism~\cite{JMMM}.} We can define T$_P$ as the
temperature at which the pseudogap,
$\Delta^{P}$(T$_P)=0$,
vanishes. The existence of particle-particle
($\Delta^P(T)>0$)
or particle-hole ($\Delta^{e-h}(T)>0$)
instability and the corresponding solution for a positive
pseudogap ($\Delta(T)>0$) can be formulated at an arbitrary $U>0$
by combining both equations (\ref{charge_gap})
 and (\ref{pair_gap})
 in one as,
\begin{eqnarray}
\Delta(T) = \left\{ \begin{array}{ll}
\Delta^{e-h}(T) & \mbox{if $U>U_c(T)$} \\
\Delta^{P}(T) & \mbox{if $U<U_c(T).$} \\
\end{array}
\right. \label{gap}
\end{eqnarray}
At zero temperature, $\Delta(0)= 0$ at $U=0$ or $U=U_c(0)$.
%%%%%%%%%%%%%%%%%%%%%%%%%%%%
\begin{figure} %[h]
\begin{center}
\includegraphics*[width=20pc,height=20pc]{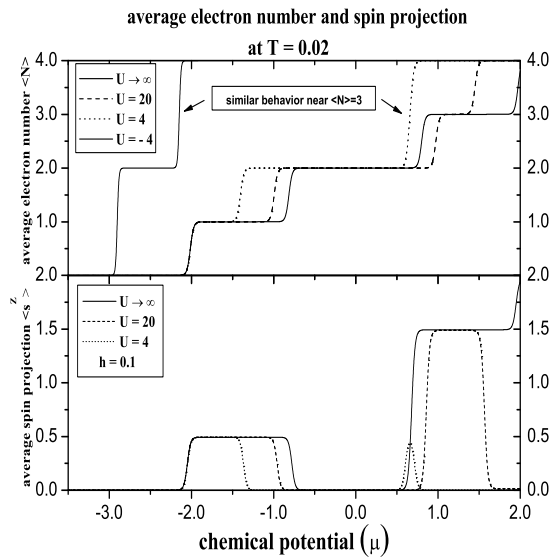}
\end{center}
\caption {{Variation of average electron concentration
$\left\langle N\right\rangle$
 (top) and  average spin
$\left\langle s^z\right\rangle$ (bottom) vs $\mu$ for various $U$
values at temperature $T=0.02$. The vanishing of the charge gap
 near $\left\langle N\right\rangle = 3$ for $U=4$
 has implications related to pairing as discussed in the text.
 Note also that the spin plot has been obtained with an applied
magnetic field of $h=0.1$ and  shows the stabilization
of a magnetic state near $\left\langle N\right\rangle = 3$
 for $U=4$. At zero field, $\left\langle
s^z\right\rangle$ = 0 everywhere due to degeneracy between spin up
and down states.}}
\label{fig:num}
\end{figure}
%%%%%%%%%%%%%%%%

\section{Results}\label{Results}

\subsection{$\left\langle N\right\rangle$ and
$\left\langle s^z\right\rangle$ vs $\mu$ and pseudogaps}\label{A}
In Fig.~\ref{fig:num}, we explicitly illustrate the variation of
$\left\langle N\right\rangle\leq 4$ {\it vs} $\mu$ for various
$U$ values in order to track the variation of charge gaps with
$U$. The opening of such gaps is a local correlation effect and
clearly does not follow from long range order, as exemplified
here. The true gaps at $\left\langle N\right\rangle=1$ and
$\left\langle N\right\rangle=4$ develop for infinitesimal $U>0$
and increase monotonically. In contrast, the charge gap at
$\left\langle N\right\rangle=3$ opens at finite $U>U_c(T)$ (see
Fig.~\ref{fig:num}). Thus at low temperature, $\left\langle
N\right\rangle$ (expressed as a function of $\mu$ in
Fig.~\ref{fig:num}) evolves smoothly for $U\leq U_c(T)$, showing
finite leaps across the MH plateaus only at $\left\langle
N\right\rangle=1$, $\left\langle N\right\rangle=2$. Such a density
profile of $\left\langle N\right\rangle$ {\it vs} $\mu$ near
$\left\langle N\right\rangle=3$ closely resembles the one
calculated in Fig.~\ref{fig:num} for the {\sl attractive} 4-site
Hubbard cluster with $U=-4$ at $T=0.02$ and is indicative of a
possible particle pairing instability. At larger $U>U_c(T)$, an
electron-hole gap is opened at $\left\langle N\right\rangle=3$.
Therefore the cluster at $U>U_c(0)$ behaves  as a MH like insulator at
all (allowed) integer numbers ($1\le N\le 8$) with electron charge
localized (non-Fermi liquid).
In contrast, at $U\leq U_c(0)$ the chemical potential gets pinned
upon doping in the midgap states at $\left\langle
N\right\rangle\simeq$ 3.
%%%%%%%%%%%%%%%%%%%%%%%%%%%%%%%%555555

While Fig.~\ref{fig:num} shows the magnetization at a relatively
high magnetic field, its behavior at very low temperature, $T\to
0$ and infinitesimal magnetic field, $h\to 0$ is also noteworthy.
In this case, as $U$ increases, $\left\langle N\right\rangle$ and
$\left\langle s^z\right\rangle$ {\it vs} $\mu$ near $\left\langle
N\right\rangle=3$ reveal islands of stability, due to  phase
separation (see Section \ref{C}),
for various charge ($N=2$ and $N=4$)
and spin ($s$ and $s^z$) configurations as follows. Phase A
($U\leq U_c(0)$): particle-particle $\Delta^{P}(0)>0$ and
spin-spin $\Delta^{s^z=0}(0)>0$ pairing gaps with the minimal
projection of spin $\left\langle s^z\right\rangle=0$.
%
%({\it spin pairing}).
%state.
Phase B ($U_c(0)<U<4(2+\sqrt{7})\simeq 18.583$): particle-hole
$\Delta^{e-h}(0)>0$ and spin-spin $\Delta^{s^z=1/2}(0)
>0$ pairings
with the spin $s=1/2$ and unsaturated ferromagnetism,
$\left\langle s^z\right\rangle=1/2$ ({\it triplet
pairing})~\cite{Sebold}. Phase C (large $U>4(2+\sqrt{7})$):
particle-hole $\Delta^{e-h}(0)
>0$ pairing without spin gap
($\Delta^{s^z=3/2}(0)\equiv 0$)
and maximum projection
$\left\langle s^z\right\rangle=3/2$ ({\it saturated
ferromagnetism})~\cite{Mattis}.

In Phase A for $U=4$, charge and spin are coupled ({\it i.e. no
charge-spin separation}), while in Phase B at $U=6$, the charge
and spin are partially decoupled ({\it partial charge-spin
separation}). In Phase C for $U \to \infty$, the charge and spin
are fully decoupled, when the charge gap saturates to its maximum
value, $\to 2(2-\sqrt{2})$, while the spin gap from $\left\langle
s^z\right\rangle=1/2$ to $\left\langle s^z\right\rangle=3/2$,
defined earlier in (\ref{spin_gap}), vanishes ({\it full
charge-spin separation}).
Phase A, due to strong particle-particle coupling with double
electron charge ($Q=2e$) and zero spin ($s^z=0$) with a majority of
$\left\langle N\right\rangle=2$
clusters,
becomes a good candidate for the full 'bosonization' of charge and
spin degrees of freedom and possible 'superconductivity'.
In contrast, at even numbers of electrons, such as $\left\langle
N\right\rangle\simeq 2$ in Fig.~\ref{fig:num}, there are
electron-hole $\Delta^{e-h}(0)>0$ and
spin-spin
$\Delta^{s^z=0}(0)>0$ pairings at all $U$ values, and therefore
the charge
$Q=2e$ and spin $s^z=0$ are both coupled
and there is full charge-spin
reconciliation, when the {\it singlet-triplet} spin excitation gap
at quarter filling approaches the charge gap,
$\Delta^{s^z=0}\equiv\Delta^{e-h}=2(2\sqrt{2}-1)$, as $U\to
\infty$. Exactly at half filling, $\left\langle N\right\rangle=4$,
there is partial charge-spin separation at all finite $U>U_c(0)$.
However, as $U\to \infty$ the charge
MH gap $\Delta^{e-h}(0)\to \infty$ and
the AF gap
$\Delta^{s^z=0}(0)\to 0$ (vanishes); there is full charge-spin
separation
with saturated spin $\left\langle s^z\right\rangle=2$
in this limiting case. Also,
we find that for all $U$, the
clusters with a single electron at $\left\langle
N\right\rangle=1$ is a MH like insulator:
charge gap $\Delta^{e-h}(0) \to \infty$, while the spin gap
$\Delta^{s^z=1/2}(0)\to 0$ with saturated spin $
\left\langle s^z\right\rangle=1/2$.
For any given $N$ in the charge sector,
 one can easily identify an insulator or a metallic
liquid if $\Delta^{e-h}(0)>0$ or $\Delta^{e-h}(0)\equiv 0$ respectively.
 Accordingly, it is also useful to distinguish a
spin insulator, $\Delta^{s}(0)>0$, or a spin liquid
($\Delta^{s}(0)\equiv 0$) state in the spin sector.

\begin{figure} %[h]
\begin{center}
\includegraphics*[width=20pc]{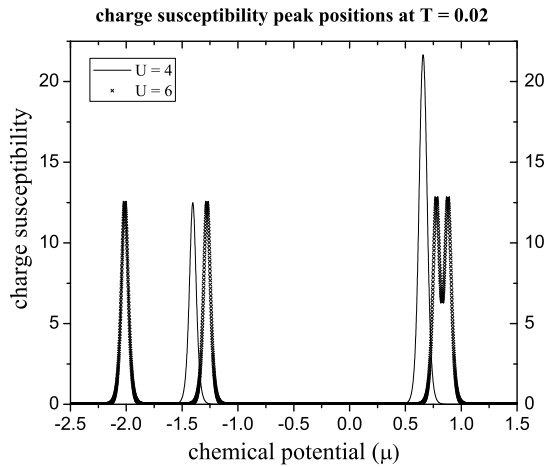}
\hfill
\end{center}
\caption {The charge susceptibility {\it vs} chemical potential
$\mu$ at  $T=0.02$ for two different $U$ values ($U=4$ and $U=6$)
below half-filling. Note that there are clearly identifiable
peaks, and such peak positions, when monitored as a function of
temperature, have been used to construct phase diagrams (see
text).}
 \label{fig:ch_sus}
\end{figure}

In Fig.~\ref{fig:ch_sus}, we show the evolution  of charge
susceptibility $\chi_c$ as a function of $\mu$, which exhibits
clearly identifiable sharp peaks. At low temperature,  peak
structures in  $\chi_c(\mu)$ and zero magnetic field spin
susceptibility, $\chi_s(\mu)$, are observed to develop in these
clusters; between two consecutive peaks, there exists a pseudogap
in charge or spin degrees. The opening of such distinct and
separated pseudogap regions for spin and charge degrees of freedom
(at low temperature) is a signature of corresponding charge and
spin separation away from half-filling.

\subsection{Pairing gap at $\left\langle N\right\rangle=3$}\label{B}

\begin{figure} %[h]
\begin{center}
\includegraphics*[width=20pc]{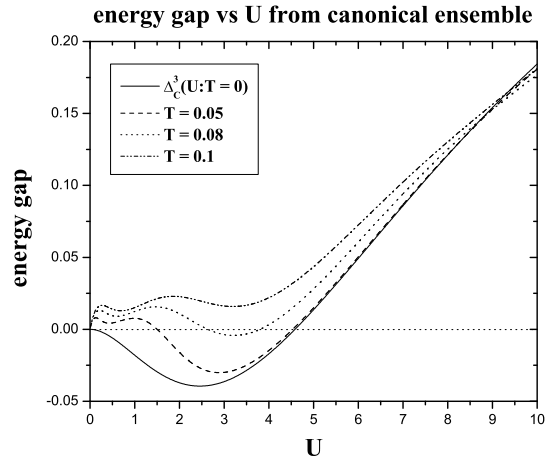}
\end{center}
\caption {The energy gap given by Eq.~(\ref{gap_3}) at zero
 temperature and its evolution at several nonzero temperatures
 as a function of $U$ obtained from the canonical ensemble.
 This is directly related to the electron-electron
 pairing gap ($U<U_c(T)$) and electron-hole charge gap ($U>U_c(T)$) as
 defined in the text. Note that charge pairing is unlikely to occur above
 temperature $T$=0.08. As previously, all energies are measured in
 units of $t$, the hopping parameter.
} \label{fig:pair}
\end{figure}

 At relatively large $U\ge U_c(T)$, the energy gap
$\Delta^{c}_{3}(U:T)=E(4;U:T)+E(2;U:T)-2E(3;U:T)$ becomes positive
for $\left\langle N\right\rangle=$ 3 (see Fig.~\ref{fig:pair}).
Its zero temperature value was first derived analytically
~\cite{JMMM} as,
\begin{eqnarray}
\Delta^c _{3} (U:T=0) = -{2\over \sqrt3}\sqrt{({16t^2}+
{U^2})}\cos{\gamma\over 3}+\nonumber\\
{U\over 3}-{2\over 3}\sqrt{({48t^2} +
{U^2})}\cos{\alpha\over 3}+\nonumber\\
\sqrt{32t^2 + U^2 + 4\sqrt{64t^4 + 3t^2U^2}}\label{gap_3},
\end{eqnarray}
where $\alpha = \arccos{\{({4Ut^2\over 3}-{U^3\over 27}) /
({16t^2\over 3} +{U^2\over 9})^{{3\over 2}}\}}$ and
$\gamma=\arccos{\{(4Ut^2)/ {({16t^2\over 3}+{U^2\over 3})^{3\over
2}}\}}$. Due to (ground state) level crossings~\cite{Mattis}, the
exact expression (\ref{gap_3}) is valid only in a limited range of
$U\leq 4(2+\sqrt{7})$.
The critical value, $U_c(0)=4.58399938$, at which
$\Delta^c_{3}(U_c: T=0)=0$, reported in Ref.~\cite{JMMM}, serves as an
estimation of the accuracy of the gap value, which is slightly
different from a value obtained in Ref.~\cite{tk}.

When $\Delta^{c}_{3}(U:T)=E(4;U:T)+E(2;U:T)-2E(3;U:T)$ becomes
negative for $U\le U_c(T)$ as shown in Fig.~\ref{fig:pair}, the
$\left\langle N\right\rangle=3$ states become less (energetically)
favorable when compared with $\left\langle N\right\rangle=2$ and
$\left\langle N\right\rangle=4$ states. This is a manifestation of
electron binding where, despite bare electron repulsion, electron
pairs experience an attraction~\cite{tk,scalettar,white}. We have
also observed a similar vanishing of charge gaps for negative $U$
values (see Fig.~\ref{fig:num}), where there is an inherent
electron-electron attraction, supporting the above statement. For
$U\geq U_c(T)$, the gap in the electron-hole channel is positive
(i.e. $\Delta^{c}_{3}(T)>0$) which favors excitonic, electron-hole
pairing, similar to MH gap at half-filling.

%%%%%%%%%%%%%%%%%%%%%%%%%%%%%%%%%%%%%%%%%%%%%%%%%%%%%%%%%%
%%%%%%%%%%%%%%%%

%%%%%%%%%%%%%%%%
\begin{figure} %[h]
\begin{center}
\includegraphics*[width=20pc]{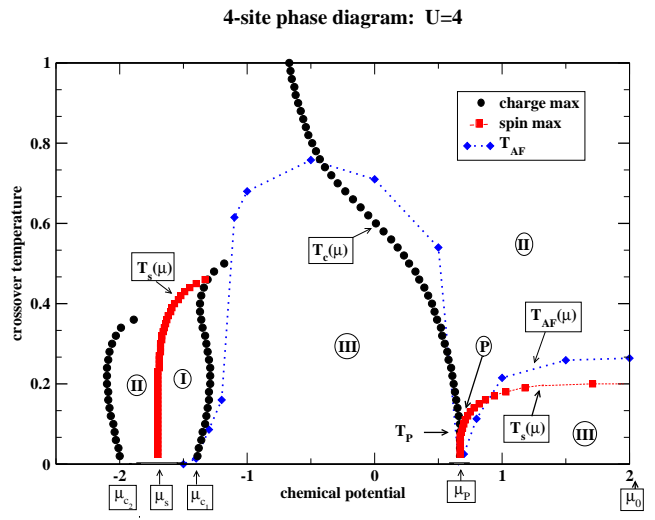}
\end{center}
\caption {Temperature $T$ vs chemical potential $\mu$ phase
diagram for the four-site cluster at $U=4$ and $h=0$, below half
filling ($\mu\leq\mu_0$). Regions I and II are paramagnetic phases
and quite similar to the ones found in the two-site cluster
(Ref.~\cite{JMMM}), showing strong charge-spin separation. Phase
III is a MH antiferromagnetic phase (with zero spin). However,
note the (new) line phase (labeled P) which consists of charge and
spin fluctuations. This is a new feature seen in the 4-site
cluster which suggests the existence of electron-electron pairing
and phase separation into hole-poor and hole-rich regions
at low temperature. For $U=4$, pairing occurs below the
temperature, T$_P$=0.076.}
\label{fig:phase_u4}
\end{figure}
%%%%%%%%%%%%%%%%%%%%%%%%%%%%55
\begin{figure} %[h]
\begin{center}
\includegraphics*[width=20pc]{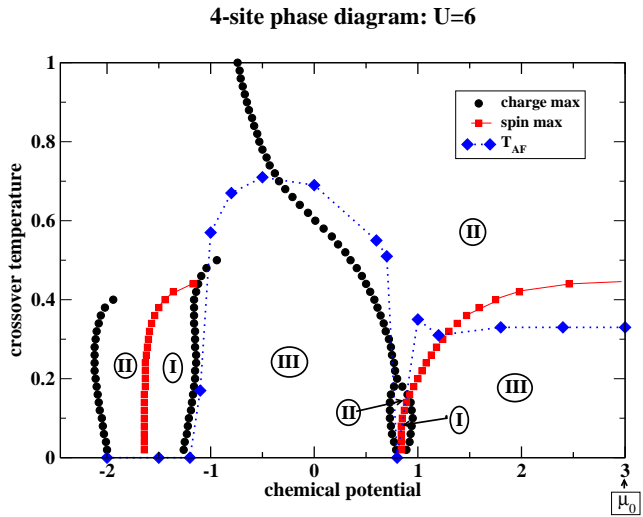}
\end{center}
\caption {Temperature $T$ vs chemical potential $\mu$ phase
 diagram for the four site cluster at $U=6$ and $h=0$, below half-filling
($\mu\leq \mu_0$). Regions I, II and III are quite similar to the
ones found for U=4 in Fig.~\ref{fig:phase_u4}, again showing
strong charge-spin separation. However, a charge gap opens as a
new bifurcation (I and II phases) which consists of charge and
spin pseudogaps, (replacing the equilibrium line phase P in the
previous figure, i.e. no electron-electron pairing here: see text
for details).  Temperature labels similar to those shown in
Fig.~\ref{fig:phase_u4} may be used here but have been suppressed
for clarity. } \label{fig:phase_u6}
\end{figure}

\subsection{Phase separation}\label{C}

It appears that the canonical approach yields an adequate
estimation of a possible pair binding instability in the ensemble of
small clusters at relatively low temperature. The competition
between attraction and repulsion under hole-doping can lead to a
`microscopic phase separation', which comprises of an
inhomogeneous structure of competing and coexisting hole-rich
$\left\langle N\right\rangle=2$ (hole-hole or electron-electron
pairing), hole-poor $\left\langle N\right\rangle=4$ (AF at half
filling) and magnetic
%%%%%%%%%%%%%%%%%%%%%%%%%%%%
$\left\langle N\right\rangle=3$
%%%%%%%%%%%%%%%%%%%%%%%%%%%
clusters.
%%%%%%%%%%%%%%%%%
From Eqs. (\ref{mupm1}) and (\ref{mupm2}), it is apparent that the
condition $\mu_{-}=\mu_{+}$
at $U=U_c(0)$ and $\left\langle N\right\rangle=3$
defines a lower bound for the existence of a phase separation
boundary that distinguishes a charge-spin separated region from a
charge-spin coupled
%region.
one.
%at which the system becomes unstable
%against local
%{\it charge}
%density fluctuations.
%%%%%%%%%%%%%%%%%

If we neglect every second hopping term in the two dimensional
square lattice, the system can be thought of as  an ensemble of
decoupled 4-site non-interacting clusters~\cite{CRG,tk}, as shown
in Fig.~\ref{fig:ensemble}. New and important features appear if
the number of electrons $\left\langle N\right\rangle=3$ and total
magnetization $\left\langle s^z\right\rangle=0$ are kept fixed for
the whole system of decoupled clusters, placed in the (particle)
bath, by allowing the particle number on each separate cluster to
fluctuate. One is tempted to think that due to symmetry, there is
only a single hole on each cluster within the $\left\langle
N\right\rangle=3$ in ensemble. However, this statement reflects a
simple average only. Due to thermal and quantum fluctuations in
the density of holes between the clusters (for $U<U_c(T)$), it is
energetically more favorable to form pairs of holes. In this case,
snapshots of the system at relatively low temperatures and at a
critical value ($\mu_P$ in Fig.~\ref{fig:phase_u4}) of the
chemical potential would reveal equal probabilities of finding
(only) clusters that are either hole-rich $\left\langle
N\right\rangle=2$ or hole-poor $\left\langle N\right\rangle=4$.
% NEW

However, at higher temperatures when  pairing coexists with
magnetic spin fluctuations, there exists a small window of
parameters which brings some stability to $\left\langle
N\right\rangle=3$ clusters
above $T_P$.
%The
Thus the
crossover from full separation (segregation) to the
coexisting magnetic ($\left\langle N\right\rangle=3$) and
hole-rich ($\left\langle N\right\rangle=2$) phases develops
smoothly and depends on the degree of disorder, i.e. temperature.
In section~\ref{A}, it was shown how the changes in the parameter
$U$ affect the spin (magnetization) by changing the cluster
configuration.

Phase separation of magnetic
%%%%%%
($\left\langle N\right\rangle=3$)
%%%%%%
and paired
%%%%%%
($\left\langle N\right\rangle=2$) states
%%%%%%
can also be triggered by increasing the magnetic field. The phase
separation
({\it i.e. segregation})
into separate magnetic $\left\langle
N\right\rangle=3$ and hole-rich $\left\langle N\right\rangle=2$
regions seen here
%%%%%%
at $\mu_{-}>\mu_{+}$,
%%%%%%
closely resembles the phase separation
effect detected recently in super-oxygenated
La$_{2-x}$Sr$_x$CuO$_{4+y}$, with various Sr
contents~\cite{Hashini}. Thus our results are consistent with the
experimental observation that  a small (applied) magnetic field mimics
doping and promotes stability of the magnetic phase
$\left\langle N\right\rangle=3$ over the superconducting state
$\left\langle N\right\rangle=2$ near optimal doping.

In addition, our calculated  probabilities (from the grand
canonical ensemble) at low temperature show that, in the interval
$\mu_{+}<\mu<\mu_P$, $\left\langle N\right\rangle=2$ clusters
become the majority; i.e. we observe both, pairing and phase
separation at low temperature.
%%%%%%%%%%%%%%%%%%%%%%%%%%%%%%%%%%%%%%%%%%%
Note that phase separation here has a different origin and occurs
at relatively weak coupling $U<U_c$ far from the level crossing
regime, at which the spin gap vanishes. Therefore, the mechanism
of phase separation here is different from the one found in
Refs.~\cite{Visscher,Emery} at large $U$ limit due to the {\it
spin} density fluctuations ($h_{+}<h_{-}$) and {\it spin}
saturation. Indeed, the four-site cluster at large $U >
4(2+\sqrt{7})$ reveals ferromagnetism in accordance with the
Nagaoka theorem~\cite{Mattis}.
%%%%%%%%%%%%%%%%%%%%%%%%%%%%55

%%%%%%%%%%%%%%%%%%%%%%%%%%%%%%%%%%%%%%%%%%%%%%%5
\subsection{Phase diagrams}

In Figs. \ref{fig:phase_u4} and \ref{fig:phase_u6}, phase diagrams
for the 4-site cluster ($U=4$ and $U=6$) are shown (see also
Ref~\cite{cond-mat}). These have been constructed almost
exclusively using the temperatures, $T_c(\mu)$, $T_s(\mu)$ and
$T_{AF}(\mu)$, defined previously. We have identified the
following phases in these diagrams: (I) and (II) are MH like
paramagnetic phases with a charge pseudogap separated by a phase
boundary where the spin susceptibility reaches a maximum, with
$\Delta^{e-h}(T)>0$, $\Delta^{AF}\equiv 0$; at finite temperature,
phase I has a higher $\left\langle N\right\rangle$ compared to
phase II; Phase (III) is a MH like antiferromagnetic insulator
with bound charge and spin, when $\Delta^{e-h}(T)>0$,
$\Delta^s(T)>0$, $\Delta^{AF}(T)>0$; (P) is a line phase for $U=4$
with a vanished charge gap at $\left\langle N\right\rangle=3$, now
corresponding to the opening of a pairing gap
($\Delta^P(T)>0$)
in the electron-electron channel with $\Delta^c_{3} (U: T)<0$. We
have also verified the well known fact that the low temperature
behavior in the vicinity of half-filling, with charge and spin
pseudogap phases coexisting, represents an AF insulator
~\cite{JMMM}. However, {\sl away from half-filling}, we find very
intriguing behavior in thermodynamical charge and spin degrees of
freedom.

In both phase diagrams, we find similar paramagnetic MH (I), (II)
charge-spin separated phases in addition to the AF (III) phase
where spin and charge are bounded. In Fig.~\ref{fig:phase_u6},
spin-charge separation in phases (I) and (II) originates for
relatively large $U$ ($=6$) in the underdoped regime. In contrast,
Fig.~\ref{fig:phase_u4} shows the existence (at $U=$ 4) of a line
phase (with pairing) similar to $U<0$ case with electron
pairing
($\Delta^P(T)>0$),
when the chemical potential is pinned up on
doping within the highly degenerate midgap states near
(underdoped) $1/8$ filling.

\subsection{Quantum critical points}

Among other interesting results, rich in variety, sharp
transitions at
%quantum critical points (QCP)
QCP
near $\left\langle N\right\rangle=3$ are found in the ground
state at $U>U_c(0)$ between phases with true charge and spin gaps;
for infinitesimal $T\to 0$, these gaps are transformed into
`pseudogaps' with some nonzero weight between peaks (or maxima) in
susceptibilities monitored as a function of doping (i.e. $\mu$) as
well as $h$. In the limiting case $T_c(\mu_c)\to 0$, the QCP
doping, $\mu_c$, defines a sharp MH like (AF)
like
transition with diverging $\chi_c$~\cite{JMMM}. At the QCP doping,
$\mu_s$, with $T_s(\mu_s) \to 0$, the zero spin susceptibility,
$\chi_s$, also exhibits a maximum.

In Fig.~\ref{fig:phase_u6}, the critical temperature $T_s (\mu)$
falls abruptly to zero at the QCP doping, $\mu_s$ (true only for
$U>U_c(0)$), implying~\cite{Tallon} that the (spin) pseudogap can
exist independently of possible particle pairing in
Fig.~\ref{fig:phase_u6}. In contrast, for $U<U_c(0)$ and low
temperature, there is no charge-spin separation near $\left\langle
N\right\rangle=3$. Therefore in Fig.~\ref{fig:phase_u4} at $U=4$
($U<U_c(0)$) we do not observe any QCP associated with $\mu_s$ or
$\mu_c$ close to $\left\langle N\right\rangle=3$. Instead,
Fig.~\ref{fig:phase_u4} shows the existence  of a line phase (with
pairing) similar to the attractive $U<0$ case with a spin
pseudogap, which exists only at finite temperature $T_s(\mu)>0$,
and electron pairing
%(T$_P>T_s$)
($T_P>T_s$),
when the chemical potential is
pinned up on doping within the highly degenerate midgap states
near (underdoped) 1/8 filling.

 We have
also seen that a reasonably strong magnetic field can bring about
phase separation and has a dramatic effect (mainly) on the QCP at
$\mu_s$, at which the spin pseudogap disappears. It is evident
from our exact results that presence of QCP at zero temperature
and critical crossovers at $T_c(\mu)$, $T_s(\mu)$ and
$T_{AF}(\mu)$ temperatures, gives strong support for the cooperative
character of existing phase transitions and crossovers similar to
 those seen in large
thermodynamic systems at finite temperatures~\cite{Langer,Cyrot}.

\subsection{Charge-spin separation}

Charge-spin separation effect is fundamental for understanding of
the generic features common for small and large thermodynamic
systems. We formulated exact criteria when the charge and spin
excitations are decoupled at $U>U_c(T)$. There is controversy
regarding the nature of MH and AF transitions and relation between
their consequent critical temperatures~\cite{JMMM}. In Figs.
\ref{fig:phase_u4} and \ref{fig:phase_u6}, the charge, decoupled
from spin degree of freedom, condenses at temperatures below $T_c$
while AF spin correlations
near half-filling
are seen to develop at lower temperatures
$T_{AF}(\mu)<T_c(\mu)$~\cite{JMMM}.
However, in the limited range close to $\mu\ge \mu_c$, there is a
reverse behavior, $T_{AF}(\mu)\ge T_c(\mu)$.
Electrons were until recently thought to carry their charge and
spin degrees of freedom equally; however accurate studies of
thermodynamic response functions in nanoscale clusters show that
in real materials, these two degrees of freedom are relatively
independent from one another and can condensate at different
doping levels $\mu_c$, $\mu_s$ and transition temperatures
$T_c(\mu)$, $T_{AF}(\mu)$ and $T_{s}(\mu)$ shown in
Fig.~\ref{fig:phase_u4} and Fig.~\ref{fig:phase_u6}.

The charge-spin separation is an unusual behavior of electrons in
some materials under certain conditions permitting the formation
of two independent (bound)  electron-electron or electron-hole
pairs (quasiparticles) in charge sectors and spin singlet and
triplet states in spin sectors. The spin quasiparticle only
carries the spin degree of the electron but not the charge, while
the charge quasiparicle has spin equal to zero but its electric
charge equals either zero (electron-hole pair) or a charge of two
electrons (electron-electron pair). We find that at large
$U>U_c(T)$,  clusters with localized charge are favored over
itinerant ones.

\begin{figure} %[h]
\begin{center}
\includegraphics*[width=20pc]{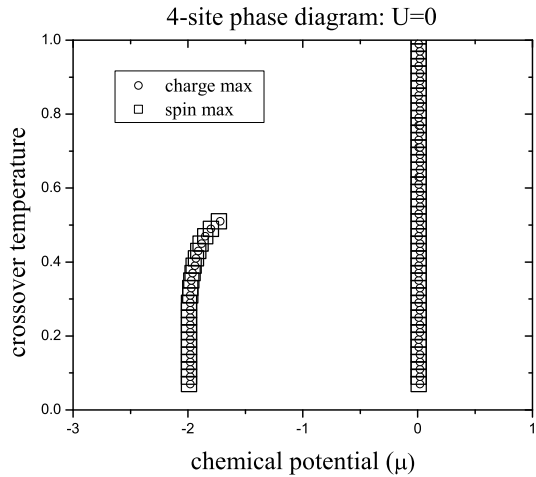}\hfill
\end{center}
\caption {The single particle or `noninteracting' ($U=0$) case,
illustrating the positions of  charge and spin susceptibility
peaks in a $T-\mu$ space for the 4-site cluster at $\mu<1$ (half-filling is
at $\mu_0=0$). Note
how the charge and spin peaks follow one another. Even in the
presence of a nonzero magnetic field there is no charge-spin
separation.}
 \label{fig:phase_u0}
\end{figure}

%\subsection{Single particle case}
As an important footnote, in the noninteracting $U=0$ case shown
in Fig.~\ref{fig:phase_u0}, the charge and spin peaks follow one
another (in sharp contrast to the $U=4$  and $6$ cases). In
regions I and II, positions of charge (as well as spin) maxima and
minima coincide indicating that there is no charge-spin
separation, even in the presence of a magnetic field. In the
entire range of $\mu$, the charge and spin fluctuations directly
follow one another without charge-spin separation. Our detailed
analysis of the (responses such as)  variation of electron
concentration $\left\langle N\right\rangle$, zero magnetic field
magnetization $\left\langle s^z\right\rangle$ {\it vs} $\mu$ and
various fluctuations shows that there is no charge-spin separation
and both, the spin and charge degrees, closely follow each other.
Thus, at $U=0$ the spin and charge degrees are strongly coupled to
one another. On the other hand, the charge-spin separation effect
at $U\neq 0$ led to rigorous definitions of Mott-Hubbard like,
antiferromagnetic, spin pseudogaps, particle-particle pairings and
related crossovers.

\section{Conclusion}

In summary, we have illustrated how to obtain phase diagrams and
identify the presence of temperature driven crossovers, quantum
phase transitions and charge-spin separation for any $U\ge 0$ in
the four-site Hubbard {\it nanocluster} as doping (or chemical
potential) is varied. Specifically, our exact solution pointed out
an important difference between the $U=4$ ($U<U_c(0)$) and $U=6$
($U>U_c(0)$) phase diagrams at finite temperature in the vicinity
of hole doping $\approx $ 1/8 which can be tied to possible
electron-electron pairing due to overscreening of the repulsive
interaction between electrons in the former.
%%%%%%%%
The resulting phase diagram with competing hole-rich
($\left\langle N\right\rangle=2$), hole-poor ($\left\langle
N\right\rangle=4$) and magnetic ($\left\langle N\right\rangle=3$)
phases
captures also the essential features of phase separation in doped
La$_{2-x}$Sr$_x$CuO$_{4+y}$~\cite{Hashini}.
%%%%%%
Our analytical results near
$\left\langle N\right\rangle\approx 3$ strongly suggest  that
particle pairing can exist at $U<U_c(T)$, while particle-hole
binding is presumed to occur for $U>U_c(T)$. It is also clear that
short-range correlations alone are sufficient for pseudogaps to
form in small and large clusters, which can be linked to the
generic features of phase diagrams in temperature and doping
effects seen in the HTSCs. The exact cluster solution shows how
charge and spin
(pseudo)
gaps are formed at the microscopic level and their behavior as a
function of doping (i.e. chemical potential), magnetic field and
temperature. The pseudogap formation can also be associated with
the condensation of spin and charge degrees of freedom below spin
and charge crossover temperatures. In addition,
our exact analytical and
numerical calculations provide important benchmarks for comparison
with Monte Carlo, RSRG, DCA and other approximations.

Finally, our results
%%%%%%
on the existence of QCP and crossover temperatures
show the {\sl cooperative  nature} of phase transition phenomena in
finite-size clusters similar to large thermodynamic systems. The
small {\it nanoclusters} exhibit a pairing mechanism in a limited
range of $U$, $\mu$ and $T$ and share very important intrinsic
characteristics with the HTSCs apparently because in all these
`bad' metallic high
%$T_c$
T$_c$
materials, local interactions play a key role. As charge and spin
fluctuations are relevant to the charge and spin susceptibilities
(\ref{fluctuation_number}), energy fluctuations are related to the
specific heats and these new results for the 4-site and larger
clusters, which provide further support to our picture developed
here, will be reported elsewhere.

One of us (ANK) thanks Steven Kivelson for interest and helpful
discussion by communicating the results of their preprint (Ref.~\cite{tk}).
This research was
supported in part by the U.S. Department of Energy under Contract
No. DE-AC02-98CH10886.

\end{document}